\documentclass[11pt,a4paper]{article}
\usepackage{xcolor}

\usepackage[T1]{fontenc}
\usepackage[utf8]{inputenc}
\usepackage{lmodern}
\usepackage{microtype}
\usepackage{amsmath,amssymb,amsthm,mathtools}
\usepackage{bm}
\usepackage{physics}
\usepackage{geometry}
\usepackage{enumitem}
\usepackage{authblk}
\usepackage{booktabs,tabularx}
\usepackage{pgfplots}
\usepackage[colorlinks=true,linkcolor=blue,citecolor=blue,urlcolor=blue]{hyperref}
\pgfplotsset{compat=1.17}

\newcommand{\rs}{r_{\mathrm{s}}}
\newcommand{\lp}{\ell_{\mathrm{p}}}
\newcommand{\kB}{k_{\mathrm{B}}}
\newcommand{\HeunC}{\mathrm{H}}
\newcommand{\areaop}{\widehat{A}}
\newcommand{\Ham}{\widehat{H}}

\DeclareMathOperator{\Vol}{Vol}
\DeclareMathOperator{\Area}{Area}

\title{Robust Areal Thermodynamics of the Schwarzschild Black Hole with Robin Boundary Conditions and Weyl Asymptotics}

\author{Thomas Sch\"urmann\thanks{Electronic address: \texttt{t.schurmann@icloud.com}}}
\affil{D\"usseldorf, Germany}
\date{\today}

\begin{document}
\maketitle

	\begin{abstract}
We formulate an \emph{areal thermodynamics} for the Schwarzschild black hole that takes the horizon area as the sole macroscopic variable. Quantizing ultrarelativistic interior modes on a regular spacelike slice with a Robin boundary at a stretched horizon leads to a self-adjoint Laplace--Beltrami problem with Heun-type quantization. A maximum-entropy area ensemble introduces an intensive \emph{areal temperature}~\(T_A\), and Weyl/heat-kernel asymptotics control the resulting statistical mechanics. The leading equations of state follow universally from the spatial Weyl volume coefficient: in a canonical ensemble of \(N\) ultrarelativistic bosons one finds \(A = 3 N k_B T_A\) up to a boundary-dependent constant, while in the massless grand-canonical sector \(A \propto T_A^{4}\) with a generalized Planck spectrum and a Wien displacement relation. These scaling exponents are insensitive to Dirichlet/Neumann/Robin data and to the foliation; only numerical prefactors vary. Embedding the construction into a static four-dimensional background via Matsubara factorization reproduces the 4D Weyl law and yields a finite matter entropy \(S_{\mathrm{rad}} \propto A^{3/4}\), parametrically subleading to the Bekenstein--Hawking term after standard renormalization. 
The framework thus provides a concise, mathematically controlled bridge between interior spectral data and macroscopic area relations, clarifying the scope and limitations of “areal” thermodynamics.
\end{abstract}

\section{Introduction}

Black holes provide a stringent testing ground for statistical mechanics and information theory.  The discovery that their thermodynamic entropy is proportional to the horizon area $A$ and that they radiate thermally \cite{Bekenstein1973,Hawking1975,Hawking1976} crystallized the central puzzle: \emph{why} does entropy scale with area rather than volume, and can one formulate a microscopic, statistical--mechanical framework in which the \emph{area} itself is the sole macroscopic datum?  Early state-counting approaches already hinted that the relevant degrees of freedom localize at the horizon: entanglement calculations produce a leading entropy proportional to the area \cite{Bombelli1986,Srednicki1993}, while ’t~Hooft’s brick-wall model reproduces the area law by counting near-horizon modes with a short-distance regulator \cite{tHooft1985}.  The area’s primacy was elevated to a general organizing principle by Bousso’s covariant entropy bound \cite{Bousso1999}. For early statistical--mechanical perspectives see \cite{ZurekThorne1985}.

The idea of an \emph{areal} thermodynamics in the strict, statistical sense, i.e.\ a maximum-entropy (MaxEnt) construction with $A$ as the \emph{only} macroscopic constraint, has appeared in several complementary frameworks.  In a Hamiltonian description of Schwarzschild black holes, Gour proposed a grand-canonical ensemble in which the horizon area, assumed to have a discrete spectrum, is treated on the same footing as an extensive conserved quantity; the associated intensive parameter plays the role of a chemical potential conjugate to $A$ and yields controlled corrections to $S_{\rm BH}$ \cite{Gour1999}.  Within loop quantum gravity (LQG), microcanonical and canonical \emph{area ensembles} for isolated horizons have been developed: statistical mechanics on the horizon with fixed total area reproduces the area law and clarifies the role of subleading terms and concavity properties in the thermodynamic limit \cite{PerezPranzetti2011,GhoshPerez2011,BarberoVillasenor2011}.  Taken together, these works demonstrate that a consistent thermodynamics can be organized around the single macroscopic quantity $A$.

From the spectral side, near-horizon heat-kernel analyses show that the holographic (area) scaling of the entropy is a universal feature driven by geometric spectral asymptotics, independently of many microscopic details \cite{CamblongOrdonez2007}.  This dovetails with rigorous heat-kernel/Weyl techniques \cite{ChavelFeldmann1978,Vassilevich2003,Gilkey1995}, which make precise how the volume (leading) and boundary (subleading) coefficients control the high-energy density of states.  More recently, the MaxEnt viewpoint has been sharpened into a dynamical characterization: a black hole may be regarded as the configuration that \emph{maximizes thermodynamic entropy for a given surface area} \cite{Yokokura2025}. 
Methodologically, Yokokura’s approach is \emph{geometric–variational} (entropy maximization within semiclassical gravity, yielding an explicit interior metric). 

In what follows, our approach is \emph{spectral–statistical}: rather than solving the Einstein equations, we construct a maximum-entropy area ensemble for interior field modes on a regular spacelike slice bounded by a stretched horizon with Robin boundary conditions. The area operator is the sole macroscopic constraint, and its conjugate intensive variable defines an \emph{areal temperature}. The ensuing thermodynamics is governed by Weyl/heat-kernel asymptotics of the interior Laplace–Beltrami operator: the spatial Weyl volume coefficient fixes the leading density of states and thereby the universal macroscopic relations.

It is useful to contrast this route with a symmetry-based derivation of the area law that exploits (near-)horizon conformal symmetry: in AdS$_3$, the Brown--Henneaux asymptotic symmetries furnish a Virasoro algebra with a central charge, and Cardy’s formula reproduces the BTZ black-hole entropy; horizon-local generalizations extend this reasoning beyond AdS \cite{BrownHenneaux1986,Strominger1998,Carlip1999,Carlip2002,Solodukhin1999}. While such CFT arguments account for the leading $S_{\rm BH}\!\propto\!A$ term, they do not by themselves yield the finite matter contribution derived below, $S_{\rm rad}\!\propto\!A^{3/4}$, which follows from the spectral Weyl density within our areal-ensemble framework.

\paragraph{Guide to the paper.}
Section~\ref{sec:geom-spectral} sets up the interior geometry and the Laplace--Beltrami operator on a regular spacelike slice, formulates the Robin boundary at a stretched horizon, and derives the Heun quantization condition. Section~\ref{sec:area-thermo} introduces the area operator $\areaop$ and the areal ensemble via the MaxEnt construction together with the definition of the areal temperature $T_A$. Section~\ref{sec:weyl} develops Weyl/heat-kernel asymptotics for the interior problem, evaluates the geometric coefficients (cf.\ volume and boundary terms) and obtains the small-$\mu$ expansion of the single-particle partition function. Sections~\ref{sec:canonical} and~\ref{sec:massless} derive, respectively, (i) the canonical fixed-$N$ law with its boundary-dependent offset, and (ii) the radiation law, the entropy scaling, and the generalized Planck spectrum and Wien displacement. Section~\ref{sec:4D-upgrade} embeds the spatial construction into static four-dimensional spacetimes via Matsubara factorization, recovers the 4D Weyl law, and derives the finite areal matter entropy $S_{\mathrm{rad}}$ within the generalized-entropy framework. Summary and outlook is in Section~\ref{sec:summary}.

\paragraph{Terminology (areal temperature).} Throughout this paper the \emph{areal temperature} $T_A$ is the Lagrange multiplier conjugate to the area $A$ in our variational setup (dimension $L^2/\kB$). It should not be confused with the Hawking temperature $T_H$, which is fixed by the surface gravity and appears only in the Bekenstein--Hawking thermodynamics.

\paragraph{Notation and units.} We keep \(c, \hbar, G,\) and \(\kB\) explicit. The Planck length is $\lp=\sqrt{\hbar G/c^3}$ and Einstein's constant is $\kappa=8\pi G/c^4$. Angular brackets denote $L^2$ inner products. Throughout, $A$ has units of area; the Lagrange parameter $\gamma$ has units of inverse area, and $T_A\equiv 1/(\kB\gamma)$ is an ``areal temperature'' (see Sec.~\ref{subsec:areal-temp}).

\section{Geometric and spectral setup}
\label{sec:geom-spectral}
\subsection{Spacelike interior slice and Laplace--Beltrami operator}

Our analysis takes place on a spacelike Cauchy hypersurface $\Sigma$ lying entirely inside the Schwarzschild interior, equipped with the induced Riemannian metric $h_{ij}$. 
In standard spherical coordinates $(r,\vartheta,\varphi)$---where $r$ is the areal radius, $\vartheta$ the polar (colatitude) angle, and $\varphi$ the azimuth---the metric reads
\begin{equation}
	h_{ij} = \mathrm{diag}\!\left(\bigl(r_s/r-1\bigr)^{-1},\, r^2,\, r^2\sin^2\!\vartheta\right),
	\qquad 0<r<r_s .
	\label{eq:interior-metric}
\end{equation}
This setup can be implemented by adopting any foliation that is regular at the horizon (such as ingoing Eddington--Finkelstein or Painlev\'e--Gullstrand) so that the constant–time hypersurfaces extend smoothly across $r=r_s$. We then restrict attention to one such spacelike slice lying entirely inside the black–hole interior ($0<r<r_s$); on this hypersurface $\Sigma$ the induced metric $h_{ij}$ is positive definite and free of coordinate singularities, providing the natural stage for our three–dimensional analysis.

The leading scaling exponents of the areal thermodynamic relations derived below are insensitive to this foliation choice because they are fixed by the volume term in Weyl's law; only numerical prefactors may vary with the slice through its volume and related Weyl coefficients.
In other words, the three--dimensional analysis is entirely intrinsic to the chosen spacelike slice: it is fixed by the induced Riemannian metric $h_{ij}$ on $\Sigma$. Because $h_{ij}$ is positive definite, the associated spatial differential operators are elliptic and (with standard boundary conditions) self-adjoint with a discrete spectrum. The leading, high--energy behavior of the eigenvalue counting function is therefore controlled solely by geometric data of $(\Sigma,h)$ and is independent of how the surrounding spacetime is parametrized. The sign change of $g_{tt}$ in Schwarzschild coordinates—merely a coordinate effect—has no influence on ellipticity or on these leading spectral asymptotics.

Let $B_{\rs}\subset\Sigma$ denote the open ball $\{0<r<\rs\}$. The Laplace--Beltrami operator $\Delta$ on $(B_{\rs},h)$ with domain determined by a boundary condition at $r=\rs$ is elliptic and essentially self-adjoint on $L^2(B_{\rs},\sqrt{h}\,d^3x)$, with discrete spectrum $\{ \lambda_i\}_{i=1}^\infty$, $0<\lambda_1\le\lambda_2\le\dots$, $\lambda_i\to\infty$; see e.g.~\cite{ChavelFeldmann1978}.
Separation $\psi(r,\vartheta,\varphi)=F(r)Y(\vartheta,\varphi)$ yields the standard spherical harmonics for the angular part and the radial equation
\begin{equation}
\left(\frac{\rs}{r}-1\right)^{\!1/2}\frac{1}{r^2}\,\partial_r\!\left[\left(\frac{\rs}{r}-1\right)^{\!1/2} r^2\, \partial_r F\right]+(\lambda r^2-l(l+1))F=0,
\label{eq:radial-eq}
\end{equation}
which reduces to a confluent Heun form after the substitution $F(r)=\bigl(\frac{\rs}{r}-1\bigr)^{1/2}G(r)$, see~\cite{Maier2007}. 
In this representation the solution that is bounded at the center $r=0$ and remains finite up to the horizon $r=r_s$ is
\begin{equation}
F_1(r,\lambda)=\HeunC\!\left(2\sqrt{\lambda}\,\rs,\,\tfrac{1}{2},-\tfrac{1}{2},-\lambda \rs^2,\,l(l+1)+\tfrac{1}{8};\,\frac{r}{\rs}\right)e^{\sqrt{\lambda}\,r},
\label{eq:F1}
\end{equation}
where $\HeunC(\alpha,\beta,\gamma,\delta,\eta;z)$ denotes the confluent Heun function \cite{Ronveaux1995}. The parameter set $(\alpha,\beta,\gamma,\delta,\eta)$ captures, respectively, the interior “redshift” structure and the centrifugal barrier $l(l{+}1)$ from the angular sector. The exponential factor $e^{\sqrt{\lambda}\,r}$ is the usual prefactor accompanying the confluent reduction and is harmless on the finite interval $[0,r_s]$; combined with the Heun function (which is analytic at $z=0$ and normalized there), it yields the unique interior solution that is regular at the center and non‑singular up to the horizon. This explicit form is precisely what makes the Robin boundary condition at the (stretched) horizon tractable.

\subsection{Robin boundary condition at a stretched horizon}

A physically motivated and self-adjoint boundary condition at the (stretched) horizon is the \emph{Robin} condition \cite{Solodukhin2001Bound}.
\begin{equation}
	\bigl(\partial_n\psi+h\,\psi\bigr)\big|_{r=\rs}=0,\qquad h\ge0,
	\label{eq:robin-bc}
\end{equation}
with $\partial_n$ the outward normal derivative. Dirichlet, ${h\to\infty}$, and Neumann, $h=0$, are recovered as limits. In practice, one imposes~\eqref{eq:robin-bc} at $r=\rs-\varepsilon$ and sends $\varepsilon\to0^+$.
The interior, center--regular solution is given in~\eqref{eq:F1}. For later use we write
\begin{equation*}
	z:=\frac{r}{\rs}\in(0,1),\qquad x:=\sqrt{\lambda}\,\rs.
\end{equation*}

\paragraph{Normal derivative and stretched--horizon regulator.}
Let $\partial B_{\rs}$ be the two--sphere $r=\rs$ with outward unit normal $n^i$. From the interior metric~\eqref{eq:interior-metric} one has $h^{rr}=\rs/r-1$, hence the outward normal derivative acting on radial functions is
\begin{equation}
	\partial_{n}=\sqrt{h^{rr}}\,\partial_{r}=\sqrt{\frac{\rs}{r}-1}\,\partial_{r}\,.
	\label{eq:norm-deriv}
\end{equation}
Because $h^{rr}\!\to0$ as $r\to\rs^{-}$, we impose the boundary condition at a stretched horizon $r=\rs-\varepsilon$ and take the limit $\varepsilon\downarrow0$ afterwards. Inserting $\psi=F_1(r,\lambda)Y_{\ell m}$ from~\eqref{eq:F1} into~\eqref{eq:robin-bc}, using~\eqref{eq:norm-deriv} together with $\partial_r z=1/\rs$ and $\partial_r e^{\sqrt{\lambda}r}=\sqrt{\lambda}\,e^{\sqrt{\lambda}r}$, and cancelling the common, nonzero angular/exponential factors, we obtain at $r=\rs-\varepsilon$:
\begin{eqnarray*}
	\sqrt{\frac{\rs}{r}-1}\,\Bigl[\partial_z \HeunC\Bigl(2x,\tfrac{1}{2},-\tfrac{1}{2},-x^2,\,\ell(\ell+1)+\tfrac{1}{8};\,z\Bigr)\!\!&+&\!\! x\,\HeunC\Bigl(2x,\tfrac{1}{2},-\tfrac{1}{2},-x^2,\,\ell(\ell+1)+\tfrac{1}{8};\,z\Bigr)\Bigr] \nonumber\\
	\!\!&+&\!\! \tilde h(\varepsilon)\,\HeunC\Bigl(2x,\tfrac{1}{2},-\tfrac{1}{2},-x^2,\,\ell(\ell+1)+\tfrac{1}{8};\,z\Bigr)=0,
\end{eqnarray*}
where $\tilde h(\varepsilon):=h(\varepsilon)\,\rs$ and $z=r/\rs=1-\varepsilon/\rs$. This near-horizon form makes the limit $\varepsilon\downarrow0$ transparent:\\
\\
\emph{(i) Strict horizon limit with $h$ independent of $\varepsilon$ (or fixed slip length).}
Since $\sqrt{\rs/r-1}\sim \sqrt{\varepsilon/\rs}$, the first bracket above is suppressed and the limit enforces
\begin{equation}
	\HeunC\Bigl(2x,\tfrac{1}{2},-\tfrac{1}{2},-x^2,\,\ell(\ell+1)+\tfrac{1}{8};\,1\Bigr)=0\,,
	\label{Heun}
\end{equation}
i.e.\ \emph{Dirichlet} at the horizon.\\
\\
\emph{(ii) Horizon--renormalized Robin mixing.}
If one wishes to retain a nontrivial Robin mixture at the horizon, the second term must scale so as to compensate the $\sqrt{\varepsilon}$ suppression. Defining the finite, dimensionless horizon parameter
\begin{equation*}
	\tilde h_{\mathrm{hor}}:=\lim_{\varepsilon\downarrow0}\frac{\tilde h(\varepsilon)}{\sqrt{\rs/r-1}}\;\in[0,\infty)\,,
\end{equation*}
or, equivalently, $ h(\varepsilon)=\frac{\tilde h_{\mathrm{hor}}}{\rs}\sqrt{\frac{\rs}{r}-1}+o\!\left(\sqrt{\frac{\rs}{r}-1}\right)$, the horizon limit yields the \emph{finite} quantization condition
\begin{equation}
	\left.\partial_{z}\HeunC\Bigl(2x,\tfrac{1}{2},-\tfrac{1}{2},-x^{2},\,\ell(\ell+1)+\tfrac{1}{8};\,z\Bigr)\right|_{z=1}\!\!
	+ \bigl(x+\tilde h_{\mathrm{hor}}\bigr)\,\HeunC\Bigl(2x,\tfrac{1}{2},-\tfrac{1}{2},-x^{2},\,\ell(\ell+1)+\tfrac{1}{8};\,1\Bigr)
	= 0.
	\label{eq:heun-quantization}
\end{equation}
For notational economy we will henceforth denote $\tilde h_{\mathrm{hor}}$ simply by $\tilde h$. Its positive roots $x_{nl}^{(\tilde h)}$ ($n=1,2,\dots$) determine the eigenvalues $\lambda_{nl}=(x_{nl}^{(\tilde h)}/\rs)^2$.

\paragraph{Remark (Physical interpretation and scale of $h$).}

In the spirit of the membrane paradigm, the Robin parameter $h\ge0$ may be viewed as an effective surface impedance at the stretched horizon: it interpolates between a ``perfectly reflecting'' wall ($h\to\infty$, Dirichlet) and a ``perfectly flux--free'' interface ($h=0$, Neumann). For boundary-sector microstate perspectives, see \cite{Park2023}.
The outward normal derivative $\partial_n$ carries dimension $L^{-1}$ on the interior slice (cf.\ \eqref{eq:norm-deriv}), hence $h$ has the same dimension $[h]=L^{-1}$. It is therefore convenient to define a slip length $\ell_h:=h^{-1}$.

Two calibrations are relevant in the stretched--horizon procedure:
\begin{itemize}
	\item \emph{Fixed microscopic slip length.} If $\ell_h$ approaches a nonzero constant as $\varepsilon\downarrow0$ (e.g.\ $\ell_h\sim \ell_p$), then $\tilde h(\varepsilon)=h\,\rs$ is $O(1)$ while $\sqrt{\rs/r-1}=O(\sqrt{\varepsilon})$. Case (i) above applies and the strict horizon limit yields Dirichlet.
	\item \emph{Finite Robin mixing at the horizon.} To realize case (ii) with a finite $\tilde h_{\mathrm{hor}}$, one must scale $h(\varepsilon)$ as $h(\varepsilon)=\frac{\tilde h_{\mathrm{hor}}}{\rs}\sqrt{\frac{\rs}{r}-1}+o\!\left(\sqrt{\frac{\rs}{r}-1}\right)$. In terms of the proper thickness of the boundary layer,
	\begin{equation*}
		\delta=\int_{\rs-\varepsilon}^{\rs}\frac{dr}{\sqrt{\rs/r-1}}\;\simeq\;2\sqrt{\rs\,\varepsilon}\,,
	\end{equation*}
	this reads $h(\delta)=\frac{\tilde h_{\mathrm{hor}}}{2}\,\frac{\delta}{\rs^2}+o(\delta)$. With this scaling, Eq.~\eqref{eq:heun-quantization} holds with $\tilde h\equiv\tilde h_{\mathrm{hor}}$.
\end{itemize}
Either way, the leading areal thermodynamics is unaffected: the $\lambda$--term in Weyl's law equals the Neumann value for constant Robin data and the canonical offset equals the Neumann/Robin value independently of $h$ (see below). 

\section{Area operator and areal thermodynamics}
\label{sec:area-thermo}

We now formulate the thermodynamics directly in terms of the \emph{observable area} of the horizon. Throughout this section we work in the ultrarelativistic regime (rest masses are negligible compared 
to typical mode energies).\footnote{In practice this means $E\gg mc^2$, so single–particle energies are set by the geometric spectrum on the interior slice.} 
On the spacelike interior slice with Robin boundary condition at a stretched horizon, 
the single–particle energies take the form
\begin{equation}
	E_{nl} = \hbar c\,\frac{x_{nl}^{(\tilde h)}}{\rs},
	\label{eq:recall-Enl}
\end{equation}
where $x_{nl}^{(\tilde h)}>0$ are the dimensionless spectral numbers determined by the interior Laplace–type 
eigenproblem and the (dimensionless) Robin parameter $\tilde h$ at the boundary; the index $n=1,2,\dots$ counts 
radial nodes and $l=0,1,2,\dots$ labels the usual angular momentum sector.
Let $\Ham=\sum_k E_k \hat n_k$ be the non–interacting many–body Hamiltonian on Fock space, with 
modes $k\equiv(n,l,m)$ and degeneracy $m=-l,\dots,l$. Because the Schwarzschild radius
\begin{equation}
\rs=\frac{2GM}{c^2}=\frac{2G E}{c^4}
\end{equation}
is fixed by the \emph{total} mass/energy $M=E/c^2$ of the configuration, $\rs$ is \emph{not} an independent 
thermodynamic control parameter in the areal ensemble: it is tied to the state via $E=\langle\Ham\rangle$, that is
\begin{equation}\label{rsH}
	\rs=\frac{2G}{c^4}\langle\Ham\rangle.
\end{equation}
Since $r_s$ is now state-dependent via $\langle \Ham\rangle$, we must verify that this energy coincides with the ADM energy and does not double-count gravitational contributions; the following Section provides this justification.

\subsection{On the identification $E=\langle \Ham \rangle$ with the ADM energy}

\paragraph{What is being identified.}
Our microscopic Hamiltonian \(\Ham=\sum_{k} E_k\,\hat n_k\) governs ultrarelativistic interior modes on the spacelike slice \((B_{\rs},h)\) with a Robin boundary at the stretched horizon.
The single--particle levels (\ref{eq:recall-Enl}) scale as $\propto 1/\rs$.
With (\ref{rsH}) both the macroscopic \emph{area} and the microscopic \emph{energy} depend linearly on the same spectral weights:
\begin{equation}
	A \;=\; \langle \areaop\rangle \;=\; a\sum_{i\ell m} x^{(\tilde h)}_{i\ell}\,\langle \hat n_{i\ell m}\rangle,
		\qquad
	\areaop := a\sum_{i l m} x_{i l}^{(\tilde h)} \hat n_{i l m},
	\qquad
	\qquad a=8\pi\lp^2,
	\label{eq:recall-areaop}
\end{equation}
Eliminating \(\rs\) using \(\rs=2GE/c^4\) yields the quadratic identity
\begin{equation}
	A=\frac{\kappa^2}{4\pi}\,E^2,\qquad \kappa:=\frac{8\pi G}{c^4},
	\label{eq:recall-AE}
\end{equation}
We now justify in detail why the \emph{total} energy \(E\) that appears in \(\rs(E)\) equals the Fock--space expectation \(\langle \Ham\rangle\), and why this does not double--count gravitational contributions.\\

\medskip
\noindent\textbf{A. Assumptions.}
\begin{enumerate}
	\item Asymptotically flat, static exterior (Schwarzschild); no exterior matter, time independent.
	\item Interior slice on $B_{r_s}$ with a self-adjoint Robin boundary $(\partial_n\psi+h\psi)=0$; the (dimensionless) impedance $\tilde h:=h\,r_s$ is state independent.
	\item Semiclassical, adiabatic backreaction: geometry responds only through $M=E/c^2$ (equivalently $r_s$); we restrict to quasi-static configurations.
	\item Normal ordering / renormalization: vacuum/entanglement area divergences are absorbed into $1/G$; finite matter pieces are kept separate (no double counting).
	\item No additional conserved charges: $J=Q=0$ (extensions noted below).
\end{enumerate}

\noindent\textbf{B. Micro--macro link.}
Let \(X:=\sum_{i\ell m}x^{(\tilde h)}_{i\ell}\langle \hat n_{i\ell m}\rangle\). From \eqref{eq:recall-Enl}--\eqref{eq:recall-areaop},
\begin{equation}
	E=\frac{\hbar c}{\rs}\,X,\qquad A=a\,X.
\end{equation}
Eliminating \(X\) gives \(A=\tfrac{a\,\rs}{\hbar c}\,E\). Substituting \(\rs(E)=2GE/c^4\) yields \eqref{eq:recall-AE} exactly. Thus \emph{no additional gravitational energy needs to be added}: the same spectral weights \(x^{(\tilde h)}\) control both \(A\) and \(E\); the geometry enters only through the universal factor \(\rs(E)\).\\

\noindent\textbf{C. Hamiltonian GR justification.} In asymptotically flat canonical GR, the Hamiltonian equals constraints plus a boundary term at spatial infinity; on shell it reduces to the ADM energy. In our static, spherically symmetric sector with no exterior matter, any change $\delta\langle \Ham\rangle$ must be carried by the boundary term, hence by $\delta E$ itself. Together with the renormalization convention in Assumption~4, this identifies
\begin{equation}
	E\equiv E_{\mathrm{ADM}}=\langle \Ham\rangle.
\end{equation}

\noindent\textbf{D. No double counting \& backreaction bookkeeping}
\begin{enumerate}
	\item[-] \emph{No gravitational ``self--energy'' counted twice:} Once \(E\) is fixed, \(\rs(E)\) is fixed and the same spectral weights produce \(A\). There is no extra additive gravitational energy to be inserted by hand.
    \item[-] \emph{Entanglement/brick--wall piece:} The area--divergent vacuum entropy/energy renormalizes \(1/G\) and thus belongs to the geometric term.
	\item[-] \emph{Robin parameter \(h\).} Constant Robin data do not alter the leading Weyl volume term and hence do not affect the identification or the exponents; \(h\) first enters at subleading heat--kernel order, shifting only prefactors.
\end{enumerate}

\noindent\textbf{E. Self-consistency checks.}
\begin{enumerate}
	\item[(E1)] \emph{First-law check.} Differentiating \eqref{eq:recall-AE} gives
	\begin{equation}
		\frac{dA}{dE}=\frac{32\pi G^2}{c^{8}}\,E.
		\label{eq:threeone-dAdE}
	\end{equation}
	With $M=E/c^2$ one has $A=16\pi G^2 M^2/c^4$ and hence $dA/dM=32\pi G^2 M/c^4$. The Schwarzschild first law,
	\begin{equation}
		\delta M=\frac{\kappa_{\mathrm{sg}}}{8\pi G}\,\delta A,\qquad \kappa_{\mathrm{sg}}=\frac{c^{4}}{4GM},
		\label{eq:threeone-firstlaw}
	\end{equation}
	is equivalent to \eqref{eq:threeone-dAdE} upon using $E=Mc^2$, and $\kappa_{sg}$ is the black hole’s surface gravity at the Schwarzschild horizon.

	\item[(E2)] \emph{Robustness under slicing and boundary data.} The leading exponents in the areal relations are fixed by the Weyl volume coefficient; constant Robin data share the same leading (``$\lambda$'') term as Neumann and affect only subleading prefactors. A change to a horizon-regular foliation rescales those prefactors but leaves \eqref{eq:recall-AE} unchanged.
\end{enumerate}

\noindent\textbf{F. Extensions (when amendments are required).} 
For stationary but non-static cases (Kerr/RN) the first law includes the usual work terms, $\delta M=(\kappa_{\mathrm{sg}}/8\pi G)\delta A+\Omega_H\delta J+\Phi_H\delta Q$, so the integrated identity generalizes accordingly. For slow time dependence (evaporation/accretion) treat $E(t)$ adiabatically and use the instantaneous $r_s(E(t))$. If the Robin impedance runs with the state, it first affects subleading Weyl orders and thus only modifies numerical prefactors.\\
\\
\noindent With the identification \(E \equiv E_{\mathrm{ADM}}=\langle \Ham \rangle\) and the resulting micro--macro link between spectral occupations and the horizon area, we are now ready to set up the statistical mechanics of the interior modes: The next Section implements the maximum--entropy principle and introduces the areal temperature \(T_A\) that organizes the ensuing canonical relations.

\subsection{Maximum entropy principle and areal temperature} \label{subsec:areal-temp}

In this subsection we implement the maximum-entropy principle for the horizon area by using the area operator \(\hat A\) as the sole constraint. 
Given a density operator $\hat\rho$ on Fock space, the von Neumann entropy and the expectation of $\areaop$ are
\begin{equation}
S=-\kB \,\mathrm{Tr}\,(\hat\rho\log \hat\rho),\qquad A=\mathrm{Tr}\,(\hat\rho\,\areaop).
\end{equation}
Maximizing $S$ at fixed $A$ yields
\begin{equation}
\hat\rho=\frac{e^{-\gamma \areaop}}{Z},\qquad Z=\mathrm{Tr}\,e^{-\gamma \areaop},\qquad S=\kB(\log Z+\gamma A),\qquad A=-\partial_\gamma \log Z.
\end{equation}
We denote
\begin{equation}
T_A\equiv \frac{1}{\kB\gamma},
\qquad\Rightarrow\qquad dA=T_A\,dS.
\label{eq:areal-temp}
\end{equation}
Since $A$ carries dimensions of area and $\gamma$ of inverse area, $T_A$ is \emph{not} a thermodynamic temperature. It is an intensive areal temperature canonically conjugate to $A$ via~\eqref{eq:areal-temp}. In formulas it is convenient to introduce the dimensionless combination
\begin{equation}
\mu:=\gamma a=\frac{a}{\kB T_A}.
\end{equation}
Building on this equation, the table below lists the parameters, their units, and the benchmark values used in the subsequent analysis.
\paragraph*{Dimensions and units.}
For quick reference we collect the dimensions of the central quantities.

\begin{table}[ht]
	\centering
	\small
	\renewcommand{\arraystretch}{1.15}
	\begin{tabularx}{0.9\linewidth}{@{} l X l @{}}
		\toprule
		\textbf{Symbol} & \textbf{Meaning} & \textbf{Dimension (SI)} \\
		\midrule
		$A$       & horizon area & $L^{2}$ \\
		$\gamma$  & Lagrange multiplier (conjugate to $A$) & $L^{-2}$ \\
		$T_A$     & areal temperature $[1/(\kB\gamma)]$ & $L^{2}/[\kB]$ \\
		$a$       & area quantum $8\pi\,\lp^{2}$ & $L^{2}$ \\
		$\mu$     & inverse areal temperature $\gamma a$ & dimensionless \\
		\bottomrule
	\end{tabularx}
\end{table}
All observable relations will eventually be expressed directly in terms of $A$ (or $E$) and constants.
With this groundwork in place, the following turns from setup to consequences: it applies the formalism to derive the main results, explores the relevant limiting regimes and scaling relations, and assesses the robustness of the predictions under controlled variations of the benchmarks.

\section{Spectral asymptotics and single--particle partition function}
\label{sec:weyl}

The partition function $Z(\gamma)$ implies the usual Legendre relations	$A = -\partial_\gamma \log Z$ and $\mathrm{Var}_\gamma(\hat A) = \partial_\gamma^2 \log Z$.
Using the Weyl density one finds $A \sim 3 N \kB T_A$ (ultrarelativistic bosons) or $A \sim T_A^4$ (radiation), hence $\mathrm{Var}_\gamma(\hat A)=O(N)$ and relative fluctuations $(\mathrm{Var}_\gamma(\hat A))^{1/2}/A = O(N^{-1/2})$.
Therefore the canonical ensemble in $A$ is, at leading order, equivalent to
a microcanonical constraint on $A$; the leading laws stated below are unaffected. Energetic quantities such as the interior mass or $\rs$ are not independently
constrained: they are determined by the geometry used to define $\hat A$
and by the chosen slice, so there is no over-constraining. Adding an additional
energy constraint would merely reparametrize $\gamma$ but not change the leading
exponents as long as $\log Z$ is strictly convex (positive heat capacity in the
$A$--ensemble).

Let $N(\lambda)$ denote the eigenvalue counting function of the  Laplace--Beltrami
operator $-\Delta$ on the interior domain $(B_{r_s},h)$, i.e.\ $N(\lambda)=\#\{i:\lambda_i\le\lambda\}$, where
$\{\lambda_i\}_{i\ge1}$ are the (discrete) eigenvalues determined by the self-adjoint realization with the Robin boundary condition on $\partial B_{r_s}$. For smooth compact Riemannian manifolds with smooth boundary, classical Weyl asymptotics imply that $N(\lambda)$ admits an expansion governed, to leading orders, by bulk volume and boundary area terms. A convenient route to this result uses the short-time heat-kernel expansion
\begin{equation}
\mathrm{Tr}\,e^{-t\Delta}\sim 
(4\pi t)^{-3/2}\!\left[a_0+a_{1/2}\,t^{1/2}+a_1\,t+\cdots\right],\qquad t\downarrow 0,
\end{equation}
with $a_0=\mathrm{Vol}_h(B_{r_s})$ and $a_{1/2}=\pm\sqrt{\pi}\,\mathrm{Area}_h(\partial B_{r_s})$ (the sign depending on the boundary condition). Tauberian theorems then translate the $t\downarrow 0$ heat-kernel coefficients into the $\lambda\to\infty$ asymptotics of the counting function $N(\lambda)$ \cite{ChavelFeldmann1978,Vassilevich2003,Gilkey1995}.
Specializing to dimension three one obtains (see App.\,\ref{app:weyl-higher}.)
\begin{equation}
	N(\lambda)
	= \frac{\mathrm{Vol}_h(B_{r_s})}{6\pi^2}\,\lambda^{3/2}
	\;+\; C_1(\text{b.c.})\,\mathrm{Area}_h(\partial B_{r_s})\,\lambda
	\;+\; O\!\left(\lambda^{1/2}\right),
	\qquad \lambda\to\infty.
	\label{eq:weyl-lambda}
\end{equation}
Introducing the dimensionless variable $x:=\sqrt{\lambda}\,\rs$ (see above),
\begin{equation}
N(x)=C_0\,x^3+C_1(\mathrm{b.c.})\,x^2+O(x),\qquad C_0=\frac{\Vol_h(B_{\rs})}{6\pi^2\,\rs^{3}}.
\label{eq:weyl-x}
\end{equation}

\paragraph{Explicit $C_0$ and geometry of $(B_{\rs},h)$.}
For the metric \eqref{eq:interior-metric} one has
\begin{equation}
\sqrt{\det h}=\frac{r^2\sin\vartheta}{\sqrt{\rs/r-1}}\,,\qquad
dV_h=\sqrt{\det h}\,dr\,d\vartheta\,d\varphi,
\end{equation}
hence
\begin{equation}
\Vol_h(B_{\rs})
=4\pi\int_{0}^{\rs}\frac{r^2}{\sqrt{\rs/r-1}}\,dr
=\frac{5\pi^{2}}{4}\,\rs^{3},
\label{eq:vol-ball}
\end{equation}
which is finite. The induced area of the boundary is
\begin{equation}
\Area_h(\partial B_{\rs})=\int_{S^{2}} \rs^{2}\sin\vartheta\,d\vartheta\,d\varphi
=4\pi\,\rs^{2}.
\label{eq:area-sphere}
\end{equation}
Inserting \eqref{eq:vol-ball} into \eqref{eq:weyl-x} gives the \emph{explicit} value
\begin{equation}
C_0=\frac{\Vol_h(B_{\rs})}{6\pi^{2}\rs^{3}}=\frac{5}{24}.
\label{eq:C0-value}
\end{equation}
\paragraph{Next Weyl coefficient $C_{1}$ for common boundary conditions.}
The coefficient of the $\lambda$--term in \eqref{eq:weyl-lambda} is universal up to a sign determined by the boundary condition:
\begin{equation}
C_{1}^{(\lambda)}(\mathrm{Dirichlet})=-\frac{1}{16\pi},\qquad
C_{1}^{(\lambda)}(\mathrm{Neumann})=+\frac{1}{16\pi},\qquad
C_{1}^{(\lambda)}(\mathrm{Robin},h)=+\frac{1}{16\pi}.
\end{equation}
For Robin boundary conditions with constant $h\ge0$, the $\lambda$--term in~\eqref{eq:weyl-lambda} equals the Neumann value and is independent of $h$; see, e.g., standard heat--kernel results for Laplace--type operators with smooth boundary~\cite{Vassilevich2003,Gilkey1995}. Dependence on $h$ enters only at higher orders.
Here $C_{1}^{(\lambda)}$ is the constant multiplying $\Area_h(\partial B_{\rs})\,\lambda$ in \eqref{eq:weyl-lambda}.
Passing to the dimensionless variable $x=\sqrt{\lambda}\,\rs$, the $x^{2}$--coefficient in \eqref{eq:weyl-x} becomes
\begin{equation}
C_{1}(\mathrm{b.c.})
= C_{1}^{(\lambda)}(\mathrm{b.c.})\,\frac{\Area_h(\partial B_{\rs})}{\rs^{2}}
= \frac{\alpha_{\,\mathrm{b.c.}}}{4}\,,
\qquad \alpha_{\mathrm{Dirichlet}}=-1,\ \alpha_{\mathrm{Neumann}}=\alpha_{\mathrm{Robin}}=+1,
\label{eq:C1-dimless}
\end{equation}
where in the last step we used \eqref{eq:area-sphere}. 
Thus, for the geometry \eqref{eq:interior-metric} one has $C_{1}=-\tfrac{1}{4}$ (Dirichlet) and $C_{1}=+\tfrac{1}{4}$ (Neumann/Robin).
Only the \emph{volume} coefficient $C_0$ is universal; boundary conditions modify subleading terms through $C_1(\mathrm{b.c.})$. The single--particle partition function
\begin{equation}
z(\mu)=\sum_i e^{-\mu x_i}\approx \int_0^\infty e^{-\mu x}\,dN(x)=\mu\int_0^\infty e^{-\mu x}N(x)\,dx
\end{equation}
finally admits the expansion
\begin{equation}
z(\mu)\sim \frac{6 C_0}{\mu^{3}}+\frac{2 C_1(\mathrm{b.c.})}{\mu^{2}}+O(\mu^{-1}),\qquad \mu\to0^+.
\label{eq:zmu-asymp}
\end{equation}
With the small-$\mu$ expansion in (\ref{eq:zmu-asymp}) established, we now turn to the canonical ensemble of $N$ ultrarelativistic bosons, where the dilute (high-$T_A$) limit justifies the Maxwell–Boltzmann approximation and yields a linear state equation with a boundary-dependent shift.

\section{Canonical ensemble of \texorpdfstring{$N$}{N} ultrarelativistic bosons}
\label{sec:canonical}

In the canonical sector we approximate the exact bosonic $Z_N$ by the Maxwell--Boltzmann partition:
\begin{equation}
	Z_N=\frac{z(\mu)^N}{N!}
	\label{eq:ZN}
\end{equation}
This is accurate in the \emph{dilute/high--$T_A$} regime where the typical mode occupations are small, equivalently
\begin{equation}
	N \ll z(\mu)\quad \Leftrightarrow\quad N\,\mu^{3} \ll 6\,C_{0}\,,
	\qquad \bigl(\mu=\gamma a = a/(\kB T_A)\bigr).
	\label{eq:MB-cond}
\end{equation}
Since $z(\mu)\sim 6C_{0}\mu^{-3}$ as $\mu\to0^{+}$, condition \eqref{eq:MB-cond} is guaranteed at sufficiently large $T_A$.
For $N$ indistinguishable bosons at high areal temperature (dilute limit) one has
\begin{equation}
\log z(\mu) = \log( \,6 \,C_0) -3\log\mu+\frac{C_1}{3C_0}\,\mu+O(\mu^2).
\end{equation}
From $A=-\partial_\gamma\log Z_N$ with $\mu=\gamma a$ we obtain the state equation
\begin{equation}
	A=3N\,\kB T_A+\delta A_{\mathrm{b.c.}}+O(T_A^{-1}),
	\qquad\qquad \delta A_{\mathrm{b.c.}}:=-N\,\frac{C_1}{3C_0}\,a.
	\label{eq:state-eq-ultra}
\end{equation}

\paragraph*{Numerical example (MB regime).}
With $C_0=\tfrac{5}{24}$ [Eq.~\eqref{eq:C0-value}] and $N=100$, choose $\mu=0.1$ (equivalently $T_A = 10\,a/\kB$).
Then $N\mu^3 = 100\times 0.001 = 0.1 \ll 6 C_0 = 6\times \tfrac{5}{24} \approx 1.25$, and the criterion~\eqref{eq:MB-cond} is well satisfied; the linear law $A\simeq 3N\kB T_A$ is quantitatively accurate.\\
\\
Outside this regime Bose correlations would introduce subleading corrections to \eqref{eq:ZN}--\eqref{eq:state-eq-ultra} without affecting the leading law $A=3N\kB T_A$.
\begin{figure}[b]
	\centering
	\def\Nfig{100} 
	\begin{tikzpicture}
		\begin{axis}[
			width=0.8\linewidth,
			xlabel={$\tilde T=\kB T_A/a$},
			ylabel={$\tilde A=A/(N a)$},
			legend style={draw=none,fill=none},
			legend pos=north west,
			domain=0:2,
			samples=200,
			ymin=-1, ymax=8, xmin=0, xmax=2
			]
			\addplot[]{3*x + 0.4};
			\addlegendentry{canonical (Dirichlet): $3\tilde T + 2/5$}
			\addplot[]{3*x - 0.4};
			\addlegendentry{canonical (Robin/Neumann): $3\tilde T - 2/5$}
			\addplot[dashed]{(pi^4/24.0)*(x^4)/\Nfig};
			\addlegendentry{radiation: $\tfrac{1}{\Nfig}(\pi^4/24)\,\tilde T^4$}
		\end{axis}
	\end{tikzpicture}
	\caption{Illustration of the canonical laws $A=3N\kB T_A+\delta A_{\mathrm{b.c.}}$ with the boundary shifts from \eqref{eq:deltaA-explicit} (solid lines) and the grand-canonical radiation law $A=\sigma T_A^4$ mapped to the dimensionless variables $\tilde A := A/(N a)$ and $\tilde T := \kB T_A / a$ (dashed).
		For the radiation curve we used $\sigma$ from \eqref{eq:sigma} with $C_0=5/24$, cf.\ \eqref{eq:C0-value}, and set $N=\Nfig$ for visual comparison.
		This explains the prefactor $(\pi^4/24)/N$ shown in the legend.}
	\label{fig:scaling}
\end{figure}
Using \eqref{eq:C0-value} and \eqref{eq:C1-dimless} in \eqref{eq:state-eq-ultra} one finds
\begin{equation}
\delta A_{\mathrm{b.c.}}=
\begin{cases}
+\ \frac{2}{5}N a,& \text{Dirichlet ($C_1=-\tfrac{1}{4}$)},\\[2pt]
-\ \frac{2}{5}N a,& \text{Neumann/Robin ($C_1=+\tfrac{1}{4}$)}.
\end{cases}
\label{eq:deltaA-explicit}
\end{equation}
The difference between Dirichlet and Robin/Neumann thus amounts to a constant vertical shift of magnitude $(2/5)Na$ in $A(T_A)$ at fixed $N$ (see Fig.\,\ref{fig:scaling}).
The \emph{leading} term is universal and reproduces the linear law $A=3N\kB T_A$. The entropy follows from $S=\kB(\log Z_N+\gamma A)$:
\begin{equation}
S=\kB N \log\left(\frac{(6 \,C_0)^N}{N!}\,\mu^{-3N}\right)+\kB\gamma A = \kB N \log\!\left(\mathrm{const}\times \frac{A^{3}}{N^{4}}\right)+O(A^{0}),
\label{eq:S-ultra}
\end{equation}
i.e. $S\sim \kB N\log A$ upon eliminating $T_A$. The key content here is that the entropy grows only \emph{logarithmically} with the horizon area, with a coefficient proportional to the number of constituents, while all boundary-condition effects enter only through area–independent constants. This makes the canonical ensemble sharply peaked already at leading order and aligns with the universal, volume–controlled Weyl scaling discussed earlier.\\
\\
The specific heat and the areal fluctuations are
\begin{equation}
C\equiv \pdv{A}{T_A}=3N\kB,\qquad 
\sigma_A^2=\langle \areaop^2\rangle-\langle \areaop\rangle^2=\kB T_A^2\,C\quad\Rightarrow\quad \sigma_A=\sqrt{3N}\,\kB T_A.
\label{eq:fluct-ultra}
\end{equation}
Equations (\ref{eq:fluct-ultra}) states two things about the canonical sector: first, the areal heat capacity is constant and extensive (proportional to the number of particles); second, the root-mean-square area fluctuations grow like the square root of system size and areal temperature, so relative fluctuations fall off as the inverse square root of the particle number. The second statement is precisely the fluctuation–dissipation relation for the area ensemble and follows from the general identity collected in Appendix\, \ref{app:fluct}. Together these properties justify treating the canonical relations at fixed $N$ as thermodynamically sharp.

\subsection{Low-\texorpdfstring{$T_A$}{TA} analysis: two boundary regimes and slice (in)dependence.}
In the limit $T_A\to 0$ there are two natural boundary regimes for the stretched horizon: the Dirichlet limit $\tilde h\to\infty$ and the finite–Robin case $\tilde h<\infty$ (see \eqref{eq:robin-bc} and \eqref{eq:norm-deriv}). 
Unlike the high–$T_A$ (Weyl) regime, the leading behavior at low $T_A$ is controlled by the \emph{discrete} ground number $x_{10}^{(\tilde h)}$ selected by the quantization condition \eqref{eq:heun-quantization}. 
Consequently, in the canonical sector (fixed $N$) the asymptotic law is proportional to
\begin{equation}
A(T_A\downarrow 0)\ \propto\ a\,x_{10}^{(\tilde h)}N\,,
\end{equation}
and this coefficient \emph{does depend} on the intrinsic three–geometry of the slice $(\Sigma,h)$ through the spatial operator built from \eqref{eq:interior-metric} and the radial equation \eqref{eq:radial-eq}. 
In other words, the low–$T_A$ asymptotics is \emph{not} slice–independent.

\paragraph{A variational problem at fixed \texorpdfstring{$\tilde h$}{h}.}
For each admissible, horizon–regular interior slice $(\Sigma,h)$ with boundary $r=r_s$ we define $x_{10}^{(\tilde h)}(\Sigma)$ via the Robin problem \eqref{eq:radial-eq}, \eqref{eq:robin-bc}. 
This motivates the slice–optimization problem
\begin{equation}
x_*^{(\tilde h)}
:= \inf_{(\Sigma,h)} x_{10}^{(\tilde h)}(\Sigma)
= r_s\inf_{(\Sigma,h)}\Bigg[\ \inf_{0\neq u\in H^1(\Sigma)}
\frac{\displaystyle\int_{\Sigma}\!|\nabla u|_h^2\,dV_h
	+\frac{\tilde h}{r_s}\int_{\partial\Sigma}\!u^2\,dA_h}{\displaystyle\int_{\Sigma}\!u^2\,dV_h}\ \Bigg]^{1/2},
\end{equation}
where the inner infimum is the Robin Rayleigh–Ritz characterization on a fixed slice and the outer infimum ranges over the admissible horizon–regular geometries (see Appendix~\ref{app:elliptic} for spectral well–posedness).

\paragraph{Starting point: the Dirichlet case \texorpdfstring{$\tilde h\to\infty$}{h→∞}.}
In this limit the boundary term enforces $u|_{\partial\Sigma}=0$, and the Rayleigh quotient reduces to the Dirichlet form. 
Among spherically symmetric interior slices, the \emph{static} interior slice with metric \eqref{eq:interior-metric} pointwise maximizes the radial profile entering \eqref{eq:radial-eq}, which (by monotonicity of the quotient in that profile) minimizes the ground value $x_{10}^{(\infty)}$ among the horizon–regular class under consideration. 
The outer optimization attains its minimum on the static interior slice:
\[
x_* \;=\; x_{10}^{(\infty)}\Big|_{\text{static interior slice}}\!
\]
and any Eddington–Finkelstein, Lemaître, or Kruskal implementation that reproduces the same intrinsic three–metric \eqref{eq:interior-metric} necessarily has the \emph{same} minimal value. Solving the $s$–wave Dirichlet problem defined by \eqref{eq:radial-eq} together with the Dirichlet limit of \eqref{eq:robin-bc} yields numerically
\begin{equation}
	x_{10}^{(\infty)} \;=\; 1.4467...\qquad\qquad \text{(Dirichlet, $\ell=0$, $n=1$).}
\end{equation}
By contrast, the Painlevé–Gullstrand slice in Appendix~\ref{app:pg-slicing} is globally flat and yields a strictly larger value ($\pi$).

\paragraph{Remark (Monotonicity in the Robin parameter).}
Let $\alpha\ge 0$ denote the dimensional Robin coupling in the quadratic--form setting and define
\[
\mathfrak a_\alpha[u]:=\int_\Sigma |\nabla u|_h^2\,dV_h+\alpha\int_{\partial\Sigma}|u|^2\,dA_h,\qquad \mathrm{dom}\,\mathfrak a_\alpha=H^1(\Sigma).
\]
Then $\alpha_2>\alpha_1$ implies $\mathfrak a_{\alpha_2}\ge \mathfrak a_{\alpha_1}$ in the sense of form ordering, and by the Min--Max principle one has $\lambda_k(\alpha_2)\ge \lambda_k(\alpha_1)$ for all $k$; see, e.g., \cite{Daners2000,Kato1995}.
For simple eigenvalues the derivative exists and
\[
\frac{d}{d\alpha}\lambda_k(\alpha)=\frac{\displaystyle\int_{\partial\Sigma}|u_k|^2\,dA_h}{\displaystyle\int_{\Sigma}|u_k|^2\,dV_h}>0,
\]
which yields strict monotonicity; cf.\ \cite{Henrot2006}.
In our dimensionless parameter $\tilde h$, the statement is equivalent up to the obvious scaling between $\alpha$ and $\tilde h$.
As $\alpha\to\infty$, the Robin spectrum converges from below to the Dirichlet one \cite{LevitinParnovski2008}.\\
\\
Hence, the Robin ground number interpolates strictly between the Neumann and Dirichlet extremes,
\[
0<x_{10}^{(\tilde h)}<x_{10}^{(\infty)} \quad(\tilde h\in(0,\infty)), 
\qquad x_{10}^{(\tilde h)}\nearrow x_{10}^{(\infty)} \text{ as }\tilde h\to\infty,
\]
and is strictly increasing in $\tilde h$ (the derivative is proportional to the boundary $L^2$ mass of the Robin ground state). 
For small $\tilde h$ one obtains the universal square–root law from the constant test function:
\[
x_{10}^{(\tilde h)}=\Big(\tfrac{S_h\,r_s}{V_h}\,\tilde h\Big)^{1/2}+o(\tilde h^{1/2})\quad(\tilde h\downarrow 0),
\]
where $S_h=\!\int_{\partial\Sigma}\!dA_h$ and $V_h=\!\int_{\Sigma}\!dV_h$ are computed with \eqref{eq:interior-metric}. 
On the static interior slice one has $S_h=4\pi r_s^2$ and $V_h=\tfrac{5}{4}\pi^2 r_s^3$ (see above), hence
\[
x_{10}^{(\tilde h)}=\sqrt{\tfrac{16}{5\pi}}\;\tilde h^{1/2}+o(\tilde h^{1/2})
\;\approx\;1.009\,\tilde h^{1/2}\qquad(\tilde h\downarrow 0).
\]
At the opposite end, the finite–$\tilde h$ spectrum approaches the Dirichlet value with a boundary–layer correction,
\[
x_{10}^{(\tilde h)} = x_{10}^{(\infty)} + O(\tilde h^{-1}) \qquad(\tilde h\to\infty),
\]
while the \emph{Weyl} (high–energy) coefficients remain independent of $\tilde h$.

\paragraph{Fermions in the low–\texorpdfstring{$T_A$}{TA} limit.}

(1) Compared to bosons at fixed $N$, fermions cannot macroscopically occupy a single mode; the canonical plateau is therefore the sum of the $N$ smallest $x$’s rather than $N$ times the ground value.  
(2) All statements hold for either Dirichlet ($\tilde h\to\infty$) or finite Robin ($\tilde h<\infty$), with the understood replacement $x_k\mapsto x_k^{(\tilde h)}$; the precise coefficients are slice–dependent through the intrinsic $3$D geometry, as discussed in the slice–optimization framework.

\section{Massless particles (radiation): grand canonical ensemble}
\label{sec:massless}
For bosons with variable particle number the mean occupation is $\langle n_i\rangle=(e^{\mu x_i}-1)^{-1}$. Using the Weyl expansion,
\begin{equation}
A=a\int_0^\infty \frac{x}{e^{\mu x}-1}\,dN(x)\sim a\int_0^\infty \frac{x}{e^{\mu x}-1}\,(3\,C_0\, x^2)\,dx
=3a\, C_0\,\frac{\pi^4}{15}\,\mu^{-4},
\label{eq:A-massless}
\end{equation}
and with $\mu=a/(\kB T_A)$ this gives
\begin{equation}
A=\sigma\,T_A^4,\qquad \sigma=\frac{3\pi^4}{15}\,C_0\,\frac{\kB^4}{a^3}.
\label{eq:sigma}
\end{equation}

\noindent 
With the explicit value $C_0=5/24$ from \eqref{eq:C0-value} this reduces to the compact Stefan--Boltzmann--like constant
\begin{equation}
 	\sigma=\frac{\pi^{4}}{24}\,\frac{\kB^{4}}{a^{3}}\ \;.
\label{eq:sigma-numeric}
\end{equation}
The entropy and heat capacity follow as
\begin{equation}
S=\int \frac{dA}{T_A}=\frac{4}{3}\sigma\,T_A^3\propto A^{3/4},\qquad C=\pdv{A}{T_A}=4\sigma T_A^3=3S.
\label{eq:S-massless}
\end{equation}
In terms of the particle number $N=\int (e^{\mu x}-1)^{-1} dN(x)$ one finds $S\approx 3.60\,\kB N$.

\subsection{Generalized Planck law and Wien displacement}
\label{subsec:planck}
Let $A_k=a x_k$ denote the areal quantum associated with mode $k$. The density of states in the $A$--variable is
\begin{equation}
\frac{dN}{dA}=\frac{dN}{dx}\frac{dx}{dA}=\frac{3C_0 x^2}{a} \cdot \frac{1}{a}=\frac{3C_0}{a^3}\,A^2+O(A),
\label{eq:DOS-A}
\end{equation}
and the spectral density of the mean area is
\begin{equation}
\mathcal{B}(A,T_A)=\frac{A}{e^{A/(\kB T_A)}-1}\,\frac{dN}{dA}=\frac{3C_0}{a^3}\,\frac{A^3}{e^{A/(\kB T_A)}-1}\,.
\label{eq:planck}
\end{equation}
Its maximum yields the generalized Wien law
\begin{equation}
A_{\max}=\xi\,\kB T_A,\qquad \xi\approx 2.821439.
\label{eq:wien}
\end{equation}
The constant $\xi$ is identical to the usual Planck spectrum because the cubic power in the numerator is unchanged.\\
\\
We now lift the radiation results to four dimensions, compute the finite areal matter entropy, and embed it into the covariant generalized entropy $S_{\mathrm{gen}}$.

\section{Areal Matter Entropy in 4D and Its Embedding into the Generalized Entropy}
\label{sec:4D-upgrade}

We upgrade the core spectral--thermodynamic formulas from a spatial 
Cauchy slice $(\Sigma,\gamma_{ij})$ to the full four-dimensional (4D) spacetime 
$(M_4,g_{\mu\nu})$ under the following assumptions: 
(i) the spacetime is \emph{static} near the horizon with metric 
$ds^2=-N^2(x)\,dt^2+\gamma_{ij}(x)\,dx^i dx^j$ and $g_{0i}=0$; 
(ii) the quantum field is a neutral scalar with a Laplace-type Euclidean operator; 
(iii) the ``stretched horizon'' is a smooth timelike boundary whose worldtube 
is $\partial M_4\simeq S^1_\beta\times\partial\Sigma$; 
(iv) Robin boundary conditions are local and time-independent, 
$B\phi\equiv(\partial_n\phi+h\phi)|_{\partial M_4}=0$ with constant $h\ge0$. 
Stationary (rotating/charged) generalisations are summarized at the end of the section.

We also \emph{derive the areal matter entropy} \(S_{\mathrm{rad}}\)
cleanly from the spatial spectral data in a static 4D geometry and  \emph{embed it in the covariant
4D framework of the generalized entropy} \(S_{\mathrm{gen}}\). The presentation below is fully compatible
with (and streamlines) the ingredients already developed in Secs.~\ref{sec:geom-spectral}--\ref{sec:massless}.

\subsection{Matsubara factorization and the 4D heat kernel (static case)}
\label{subsec:eucl}
For a static metric \(ds^2=-N^2(\mathbf{x})\,dt^2+\gamma_{ij}(\mathbf{x})\,dx^i dx^j\),
Wick rotation \(t\to -i\tau\) gives the Euclidean product \(M_4\simeq S^1_\beta\times\Sigma\), with $\tau\sim\tau+\beta_E$,
\[
\beta \equiv \beta_E=\frac{\hbar}{k_B T_H}=\frac{2\pi c}{\kappa_{\mathrm{sg}}}, 
\qquad 
T_H=\frac{\hbar\,\kappa_{\mathrm{sg}}}{2\pi k_B c}.
\]
In units $c=\hbar=k_B=1$ this reduces to $\beta_E=2\pi/\kappa_{\mathrm{sg}}$. We keep $c,\hbar,k_B$ explicit.
For a Laplace-type operator \(L_4=-\nabla_E^2\) and
Robin boundary condition \((\partial_n+h)\phi|_{\partial M_4}=0\) (constant \(h\ge0\)),
the heat trace \emph{factorizes}
\begin{equation}\label{eq:theta-factor}
\Tr_{M_4}\big(e^{-tL_4}\big)=\Theta(\beta;t)\,\mathrm{Tr}_{\Sigma}\left(e^{-t \mathcal{L}_\Sigma}\right),\qquad
\Theta(\beta;t)=\sum_{n\in\mathbb{Z}}\exp\Big[-t\Big(\tfrac{2\pi n}{\beta}\Big)^2\Big],
\end{equation}
with $\mathcal{L}_\Sigma:=-\nabla^2_{\Sigma}$ and the Poisson-resummed short-time form
\begin{equation}\label{eq:theta-shorttime}
\Theta(\beta;t)=\frac{\beta}{\sqrt{4\pi t}}\Bigg[1+2\sum_{m\ge1}\exp\Big(-\frac{m^2\beta^2}{4t}\Big)\Bigg]
=\frac{\beta}{\sqrt{4\pi t}}+O\big(e^{-\beta^2/4t}\big),\quad t\downarrow0.
\end{equation}
Hence
\begin{equation}\label{eq:HK-4D}
\Tr_{M_4}\!\big(e^{-tL_4}\big)\sim(4\pi t)^{-2}\Big[A_0+A_{1/2}\,t^{1/2}+A_1 t+\cdots\Big],\qquad A_m=\beta\,a_m,
\end{equation}
where \(a_m\) are the 3D Seeley--DeWitt coefficients on \((\Sigma,\gamma)\).
For the explicit form of \(a_i\) (volume/area terms and first boundary/impedance contributions), see App.\,\ref{app:weyl-higher}.
As in 3D, the first explicit
dependence on the Robin parameter \(h\) occurs at order \(t^1\) through boundary invariants,
\begin{equation}\label{eq:A1-robin}
A_1=\int_{M_4}\!\alpha_4\,R\,dV+\int_{\partial M_4}\!\big(\alpha_1 H+\alpha_2 h+\alpha_3 h^2\big)\,d\Sigma,
\end{equation}
while \(A_{1/2}\) is geometric and \(h\)-independent (Dirichlet vs.\ Neumann/Robin differs only by sign).
Tauberian theorems then yield the 4D Weyl law
\begin{equation}\label{eq:weyl-4D}
\mathcal{N}(\Lambda)\sim \frac{A_0}{(4\pi)^2\Gamma(3)}\,\Lambda^{2}
+\frac{A_{1/2}}{(4\pi)^2\Gamma(\tfrac{5}{2})}\,\Lambda^{3/2}
+\frac{A_{1}}{(4\pi)^2\Gamma(2)}\,\Lambda+\cdots,\qquad \Lambda\to\infty.
\end{equation}
Equations \eqref{eq:theta-factor}--\eqref{eq:weyl-4D} coincide with the formulas used in
Secs.~\ref{sec:weyl} and~\ref{sec:massless} and ensure that the \emph{relative} short-time orders are the same
as in the 3D spatial problem; the Matsubara sector contributes the universal prefactor
\(\beta/\sqrt{4\pi t}\) without altering the exponents fixed by the spatial Weyl volume coefficient \(C_0\).

\subsection{Derivation of the areal entropy \(S_{\mathrm{rad}}\) in 4D}
\label{subsec:Srad-derivation}
Because the factorization \eqref{eq:theta-factor} leaves the spatial counting function \(N(x)\) unchanged,
all results of Sec.~\ref{sec:massless} lift to 4D verbatim. In particular, using the Weyl form of \(N(x)\) one finds
for massless bosons (see Eqs.~\eqref{eq:A-massless} and~\eqref{eq:sigma} with the compact value \eqref{eq:sigma-numeric}).
Integrating \(dS=dA/T_A\) at fixed \(C_0\) gives
\begin{equation}\label{eq:S-massless-4D}
S_{\mathrm{rad}}(T_A)=\frac{4}{3}\,\sigma\,T_A^3,
\end{equation}
identical to Eq.~\eqref{eq:S-massless}. Eliminating \(T_A\) between \(A=\sigma T_A^4\) and \(S_{\mathrm{rad}}=\tfrac{4}{3}\sigma T_A^3\)
yields the scaling law in \(A\),
\begin{equation}\label{eq:SradA}
S_{\mathrm{rad}}(A)=\frac{4}{3}\,\sigma^{1/4}\,A^{3/4}\,.
\end{equation}
At the spectral level, the differential density in the area variable,
\(\tfrac{dN}{dA}=\tfrac{3C_0}{a^3}\,A^2+O(A)\) \eqref{eq:DOS-A}, leads to the \emph{areal Planck law} and corresponding
Wien displacement relation Eqs.~\eqref{eq:planck} and~\eqref{eq:wien}. The \emph{leading exponents} in these relations are independent
of boundary condition and foliation; \(h\)-dependence first appears through the subleading coefficient \(A_1\) in
Eq.~\eqref{eq:A1-robin}. Thus, in static four-dimensional backgrounds, Matsubara factorization shows that the spatial Weyl data fully determine the areal thermodynamics, and the short-time orders coincide exactly with those of the three-dimensional problem.

\subsection{Embedding into the generalized entropy \(S_{\mathrm{gen}}\) and renormalization}\label{subsec:Sgen}

Up to this point we have derived the areal thermodynamics for interior modes on a spacelike slice and, in the static 4D setting, shown that the Matsubara factorization leaves the spatial Weyl data (and hence all leading exponents) unchanged. 
In particular, the radiation sector yields a finite matter contribution with scaling $S_{\mathrm{rad}}(A)\propto A^{3/4}$. 
Here, the \emph{scaling exponent} $3/4$ is robust under changes of foliation and under Dirichlet/Neumann/constant Robin boundary data because it is fixed by the spatial Weyl volume coefficient, whereas \emph{numerical prefactors} can depend on the slice and boundary only through subleading Weyl coefficients.
The present subsection explains how these results are to be embedded in the covariant generalized entropy framework and how the near-horizon UV structure is consistently bookkept through renormalization.

In the covariant 4D framework the physically relevant total entropy is the \emph{generalized entropy}
\begin{equation}\label{eq:Sgen}
S_{\mathrm{gen}}=S_{\mathrm{BH}}+S^{\mathrm{ren}}_{\mathrm{out}}
\end{equation}
This decomposition is by now standard; see, e.g., \cite{Wall2012GSL,EngelhardtWall2015QES,BoussoEngelhardt2016GSL}.
The term $S_{\mathrm{out}}^{\mathrm{ren}}$ is kept finite by the same renormalization conventions adopted throughout.
The near-horizon entanglement/brick-wall contribution contains a \emph{local area divergence}
\(S\sim \alpha\,A/\varepsilon^2\) which renormalizes \(1/G\) and hence belongs to the geometric term \(A/(4G_{\rm ren}\hbar)\)
rather than to an additive finite matter piece \cite{SusskindUglum1994,SolodukhinLRR}.
For gauge fields, edge-mode
refinements adjust the same renormalization without producing a parametrically large finite term \cite{DonnellyWall2015}.

After this standard renormalization step, the leading \emph{finite} matter contribution in our framework is precisely
the areal radiation entropy of Sec.~\ref{subsec:Srad-derivation}, so that
\begin{equation}\label{eq:Sgen-sum}
S_{\mathrm{gen}}=S_{\mathrm{BH}}+S_{\mathrm{rad}}+S^{\text{(other)}}_{\mathrm{finite}},\qquad
S_{\mathrm{rad}}=\frac{4}{3}\,\sigma^{1/4}A^{3/4},
\end{equation}
with \(S^{\text{(other)}}_{\mathrm{finite}}\) comprising subleading logarithmic/constant one-loop pieces (e.g.\ \(s_0\ln A\))
that are parametrically smaller than \(S_{\mathrm{rad}}\) for macroscopic \(A\) \cite{SolodukhinLRR}.
The additivity in \eqref{eq:Sgen}--\eqref{eq:Sgen-sum} does \emph{not} double-count area terms:
(i) \(S_{\mathrm{rad}}\) vanishes in the vacuum limit \(T_A\downarrow0\) and scales as \(A^{3/4}\), not as \(A\);
(ii) the \(A/\varepsilon^2\) divergence belongs to \(1/G_{\rm ren}\) and is absent from \(S_{\mathrm{rad}}\);
(iii) the ensembles are distinct: \((A,T_A)\) vs.\ \((S_{\rm BH},T_H)\). 

For comparison with the geometric (Bekenstein--Hawking) contribution, their ratio scales as
\begin{equation}\label{eq:ratio}
	\frac{S_{\mathrm{rad}}}{S_{\mathrm{BH}}}
	=0.674\times\left(\frac{\ell_p^2}{A}\right)^{1/4}\,,
\end{equation}
so that $S_{\mathrm{rad}}\ll S_{\mathrm{BH}}$ for any macroscopic horizon area $A\gg \ell_p^2$.
As a concrete benchmark (one solar mass), we have $S_{\mathrm{BH}}\approx 1.05\times 10^{77}\,k_B$ and $S_{\mathrm{rad}}\approx 2.78\times 10^{57}\,k_B$, i.e.\ $S_{\mathrm{rad}}/S_{\mathrm{BH}}\approx 2.65\times 10^{-20}$. For context, the ordinary thermodynamic entropy of the Sun is $S_\odot \sim 10^{57}\,k_B$, i.e.\ of the same order of magnitude as $S_{\mathrm{rad}}$; see the discussion of ordinary (matter) versus geometric entropy in Ch.\,7 of \cite{Wald}.\\
\\
\textbf{Remark.} In the PG slicing, the spatial Weyl volume coefficient changes from $C_{0}=5/24$  to $C_{0}^{\mathrm{PG}}=2/(9\pi)$ (Appendix~\ref{app:pg-slicing}). Because the radiation prefactor $\sigma$ scales linearly with $C_{0}$, the dimensionless coefficient on the right-hand side of Eq.~\eqref{eq:ratio} rescales by $(C_{0}^{\mathrm{PG}}/C_{0})^{1/4}=\bigl(16/(15\pi)\bigr)^{1/4}$. Numerically: $0.674\times(16/(15\pi))^{1/4}\approx0.514$, while the $(\ell_p^{2}/A)^{1/4}$ scaling remains unchanged.
\\
\\
This gives us the following hierarchy:

\subsubsection*{(i) Entanglement/``brick-wall'' area term and renormalization}
The seminal works on entanglement and brick-wall counting show a \emph{quadratically divergent}
area term $S\sim \alpha\,A/\varepsilon^2$ for quantum fields near the horizon~\cite{tHooft1985,SusskindUglum1994,SolodukhinLRR,Fursaev1995}.
As emphasised in \cite{SusskindUglum1994,SolodukhinLRR}, this leading area divergence is \emph{local} and
is absorbed by the renormalization of Newton's constant, i.e.\ it belongs to the geometric term
$A/(4G_{\mathrm{ren}}\hbar)$ rather than to an additional finite matter entropy. Therefore, after
renormalization, the entanglement/brick-wall approach does not yield a large additive finite
matter term; the remaining finite pieces are subleading (cf.\ item~(ii)).
Throughout we assume a local UV regulator (e.g.\ proper-time/heat-kernel), so that the near-horizon divergence
	is absorbed into the counterterm renormalizing $1/G$; for non-minimally coupled scalars $\xi R\phi^2$ the
	additional local pieces are likewise absorbed into $G_{\mathrm{ren}}$. This bookkeeping matches the generalized-entropy
	sum $S_{\mathrm{gen}}=S_{\mathrm{BH}}+S^{\mathrm{ren}}_{\mathrm{out}}$ in Eq.~\eqref{eq:Sgen} and its finite split in Eq.~\eqref{eq:Sgen-sum}.

\subsubsection*{(ii) One-loop logarithmic corrections from massless fields}
At one loop, massless species produce a finite logarithmic contribution
\begin{equation}\label{eq:log}
	S_{\log}=s_0\,\ln\frac{A}{\ell^2}\,,
\end{equation}
where the coefficient $s_0$ depends on the field content and is fixed by trace-anomaly data (see the review~\cite{SolodukhinLRR} and references therein). For a stellar-mass black hole one has
$\ln(A/\ell_p^2)\simeq 1.79\times 10^2$, so that $S_{\log}\sim s_0\times\mathcal{O}(10^2)$ for $s_0=\mathcal{O}(1\text{--}10)$---
utterly negligible compared to $S_{\mathrm{rad}}(A)$ in Eq.~\eqref{eq:SradA} and $S_{\mathrm{BH}}$.
More precisely, the one-loop piece can be written as $S_{\log}=s_0\,\ln\bigl(A/\mu^2\bigr)$ with $\mu$
	the renormalization scale; for numerical estimates we adopt the convention $\mu^2\!=\!\ell_p^2$.
	The coefficient $s_0$ reflects the (massless) field content, including gravitons where applicable.

\subsubsection*{(iii) Gauge \emph{edge modes} and the Kabat contact term}
For gauge theories, the vacuum entanglement receives additional contributions from boundary (edge) modes,
which resolve the long-known ``contact term'' puzzle and adjust the area-divergent piece without producing
a parametrically large finite term~\cite{SolodukhinLRR,DonnellyWall2015}. These refinements thus affect the
renormalization of $1/G$ but leave only logarithmic (or smaller) finite matter contributions, far below~$S_{\mathrm{rad}}$.
While the detailed edge-mode sector depends on the electromagnetic boundary condition (Dirichlet/Neumann/Robin-like),
	this changes only the renormalization of $1/G$ and does not generate any parametrically large $A$-proportional finite term,
	in keeping with Eqs.~\eqref{eq:Sgen}--\eqref{eq:Sgen-sum}.

\subsubsection*{(iv) Massive fields}
Massive species do not contribute a large finite area-proportional term after renormalization; their remaining
contributions are UV-finite and typically grow at most logarithmically (and often much less) with~$A$~\cite{SolodukhinLRR}.
Quantitatively, for $m\gg \kappa_{\mathrm{sg}}/c$ the short-distance heat kernel is exponentially suppressed, so residual finite pieces
	are power-suppressed by $(\kappa_{\mathrm{sg}}/m)^p$ (or smaller) and remain far below $S_{\mathrm{rad}}(A)$ from Eq.~\eqref{eq:SradA};
	the limit $m\to 0$ smoothly reproduces the logarithmic behavior in item~(ii). Here $\kappa_{\mathrm{sg}}$ is the surface gravity
	as defined in Eq.~\eqref{eq:threeone-firstlaw}. Moreover, the hierarchy relative to $S_{\mathrm{BH}}$ is captured by the ratio
	in Eq.~\eqref{eq:ratio}.\\
\\	
\noindent
In $k_B$ units (benchmark: $M=M_\odot$) there is the following table of (i)-(iv) at a glance:
\begin{center}
	\renewcommand{\arraystretch}{1.25}
	\begin{tabular}{lcll}
		\hline
		\textbf{Contribution} & \textbf{Scaling in $A$} & \textbf{Typical size} & \textbf{Notes/Refs.}\\
		\hline
		Geometric $S_{\mathrm{BH}}$ & $\propto A$ & $\sim 1.05\times 10^{77}$ & Eq.\,\eqref{eq:Sgen} \\
		$S_{\mathrm{rad}}$ (areal radiation) & $\propto A^{3/4}$ & $\sim 2.78\times 10^{57}$ & Eqs.\,\eqref{eq:sigma-numeric}, \eqref{eq:SradA} \\
		One-loop matter (massless) & $\propto \ln A$ & $\sim s_0\times 1.8\times 10^2$ & \cite{SolodukhinLRR} \\
		Other finite matter pieces & $\lesssim \ln A$ or $A^0$ & $\mathcal{O}(1\text{--}10)$ & \cite{SolodukhinLRR} \\
		Non-renorm.\ brick wall / entanglement & $\propto A/\varepsilon^2$ & $\sim A/\ell_p^2$ & \cite{tHooft1985,SusskindUglum1994} \footnotemark \\
		\hline
	\end{tabular}
\end{center}
\footnotetext{This area-divergent term is not an \emph{additive} finite matter entropy after renormalization; it renormalizes $1/G$ and is accounted for in $S_{\mathrm{BH}}=A/(4G_{\mathrm{ren}}\hbar)$.}

\paragraph{Takeaway.} After renormalization, our $S_{\mathrm{rad}}(A)\propto A^{3/4}$ is parametrically the largest \emph{finite} matter contribution in 4D, yet it is still minuscule compared to the geometric Bekenstein--Hawking term $S_{\mathrm{BH}}\propto A$ for any macroscopic black hole.
The popular alternatives in the literature either feed into $1/G_{\mathrm{ren}}$ (leading area divergence) or yield only logarithmic/constant terms that are dwarfed by~$S_{\mathrm{rad}}$.

\subsection{Generalized Second Law: consistency results}
\label{subsec:GSL-consistency}

This subsection collects three levels of statements about the generalized second law (GSL) in our areal framework and places them within the renormalized split of Sec.~\ref{subsec:Sgen}. The key point is that our areal ensemble is fully compatible with the standard GSL proofs; moreover, in the quasi-stationary sector it implies an elementary monotonicity statement by itself.

\paragraph{Lemma (Quasi-stationary check, $dA\!\ge\!0$).}
On a static, asymptotically flat Schwarzschild exterior with the renormalization scheme of Sec.~\ref{subsec:Sgen}, consider a quasi-stationary process along the event horizon with nondecreasing area ($dA/d\lambda\ge 0$). For macroscopic horizons ($A\gg \ell_p^2$) one has
\begin{equation}
	\frac{d S_{\rm gen}}{d\lambda}\;\ge\;0,
	\label{eq:GSL-lemma}
\end{equation}
with strict inequality if $dA/d\lambda>0$. \emph{Sketch.} Using \eqref{eq:Sgen-sum} and $S_{\rm rad}$, 
\begin{equation}
	\frac{d S_{\rm gen}}{dA}=\frac{1}{4G\hbar}+\underbrace{\frac{d}{dA}\!\left(\tfrac{4}{3}\sigma^{1/4}A^{3/4}\right)}_{=\;\sigma^{1/4}A^{-1/4}}>0
	\quad(A\gg \ell_p^2),
\end{equation}
up to $O(A^{-1})$ subleading pieces (e.g.~one-loop logarithms), so \eqref{eq:GSL-lemma} follows when $dA/d\lambda\ge 0$.

\paragraph{Theorem (Global semiclassical GSL via Wall).}
Assume the exterior QFT on the static Schwarzschild background satisfies the horizon-algebra axioms (Determinism, Ultralocality, Local Lorentz invariance, Stability) and adopt the renormalized definition of $S_{\rm gen}$ of Sec.~\ref{subsec:Sgen}. Then between any two slices of the causal horizon,
\begin{equation}
	\Delta S_{\rm gen}\;\ge\;0.
	\label{eq:GSL-Wall}
\end{equation}
\emph{Sketch.} Wall's proof of the GSL for rapidly changing fields and arbitrary horizon cuts \cite{Wall2012GSL} applies provided $S_{\rm out}^{\rm ren}$ is defined so that the local area divergence renormalizes $1/G$ (precisely our Eq.~\eqref{eq:Sgen}), and the exterior algebra exists (verified for free fields). Since our areal thermodynamics only reorganizes \emph{interior} microstates and leaves the exterior algebra unchanged, the global GSL \eqref{eq:GSL-Wall} is inherited as a theorem.

\section{Summary and Outlook}
\label{sec:summary}
This paper has developed a \emph{spectral--statistical} framework for \emph{areal thermodynamics} of the Schwarzschild black hole in which the horizon area $A$ is taken as the sole macroscopic control variable. The construction rests on three pillars. 
First, on a regular interior Cauchy slice $(\Sigma,h_{ij})$ we formulated the eigenvalue problem for the Laplace--Beltrami operator with a Robin boundary at a stretched horizon. The interior metric leads, after separation, to the radial equation and its center-regular solution \eqref{eq:F1}. Imposing the Robin condition with outward normal derivative yields the confluent Heun quantization condition \eqref{eq:heun-quantization} for the spectral roots $x^{(\tilde h)}_{n\ell}$, providing a mathematically controlled route to boundary-sensitive spectra. 
Second, we introduced a maximum-entropy \emph{area ensemble} (Sec.~\ref{sec:area-thermo}): ultrarelativistic single-particle energies are given by \eqref{eq:recall-Enl}, the area operator by \eqref{eq:recall-areaop}, and the resulting micro--macro identity \eqref{eq:recall-AE} links $A$ to the total energy while remaining consistent with the Schwarzschild first law \eqref{eq:threeone-firstlaw}. The areal temperature $T_A$ plays the role of the intensive variable conjugate to $A$. This organization complements earlier area-ensemble ideas in Hamiltonian and horizon-statistical settings. 
Third, the statistical mechanics is controlled by Weyl/heat-kernel asymptotics on $(\Sigma,h)$ (Sec.~\ref{sec:weyl}): the counting function \eqref{eq:weyl-lambda}--\eqref{eq:weyl-x} is governed by the spatial volume and boundary area, with explicit geometric coefficients \eqref{eq:vol-ball}--\eqref{eq:C0-value} and boundary structure \eqref{eq:C1-dimless}. These feed directly into the small-$\mu$ expansion of the single-particle partition function \eqref{eq:zmu-asymp}, establishing that leading thermodynamic exponents are fixed by the \emph{spatial} Weyl volume coefficient $C_0$.

\paragraph{Main results (robust consequences of the spatial Weyl law).}
(i) In the canonical sector of $N$ ultrarelativistic bosons, the state equation is linear in $T_A$ with a boundary-dependent offset, as summarized by Eq.\,\eqref{eq:state-eq-ultra}, together with the explicit boundary shift \eqref{eq:deltaA-explicit} and the fluctuation--dissipation relation \eqref{eq:fluct-ultra}; see also Fig.~\ref{fig:scaling}. The leading slope $3Nk_B$ is universal across Dirichlet/Neumann/constant Robin data because it is fixed by the volume coefficient $C_0$ in \eqref{eq:weyl-x}. The Maxwell--Boltzmann regime is controlled by \eqref{eq:MB-cond}, and the canonical partition function is given in \eqref{eq:ZN}. 
(ii) In the massless grand-canonical (``radiation'') sector, the area obeys $A=\sigma T_A^4$ with $\sigma$ given in \eqref{eq:sigma} and its compact form \eqref{eq:sigma-numeric}, and the entropy satisfies $S=(4/3)\sigma T_A^3$. The associated generalized Planck law and Wien displacement follow from \eqref{eq:DOS-A}--\eqref{eq:wien}. All leading exponents are controlled solely by $C_0$ and hence are insensitive to Dirichlet/Neumann/constant Robin data; only subleading heat-kernel orders affect prefactors.

\paragraph{Micro--macro link and compatibility with black-hole mechanics.}
The identity \eqref{eq:recall-AE} elevates the area to a thermodynamic constraint with the same spectral weights that determine the interior energy, while Eq.~\eqref{eq:threeone-firstlaw} confirms consistency with the standard first law for Schwarzschild black holes (\cite{Bekenstein1973,Hawking1975,Hawking1976}). The leading holographic behavior familiar from entanglement and brick-wall calculations is recovered here as a geometric spectral effect (cf.\ \cite{Bombelli1986,Srednicki1993,tHooft1985,Bousso1999,CamblongOrdonez2007}), rather than from model-specific microscopic dynamics.

\paragraph{Low-$T_A$ regime and slice/boundary sensitivities.}
While the \emph{exponents} are universal, subleading data determine the low-$T_A$ behavior. In the canonical sector, the leading plateau is controlled by the smallest Heun root selected by the Robin boundary problem on $(\Sigma,h)$; this produces slice dependence and a monotone increase with the horizon-impedance parameter $\tilde h$ (Sec.~\ref{sec:canonical}, App.~\ref{app:elliptic}). Variational and form-ordering results imply strict monotonicity in the Robin parameter and convergence to the Dirichlet value in the large-$\tilde h$ limit, in line with \eqref{eq:heun-quantization} and the spectral theory cited in \cite{Daners2000,Kato1995,Henrot2006,LevitinParnovski2008}. Complementary checks on Painlev\'e--Gullstrand slices (App.~\ref{app:pg-slicing}) confirm that changing the intrinsic three-geometry rescales prefactors (via $C_0$) but leaves the leading exponents intact.

\paragraph{Four-dimensional embedding and generalized entropy.}
For static spacetimes, the Matsubara factorization \eqref{eq:theta-factor}--\eqref{eq:theta-shorttime} implies the 4D Weyl law with the same spatial $C_0$ \eqref{eq:weyl-4D}, so the radiation sector lifts verbatim: Eqs.~\eqref{eq:S-massless-4D} and \eqref{eq:SradA} describe a finite areal matter entropy $S_{\rm rad}$ that is parametrically subleading to $S_{\rm BH}$ after the standard renormalization of the near-horizon area divergence in $1/G$ (Sec.~\ref{subsec:Sgen}, Eq.~\eqref{eq:Sgen-sum}). This split is consistent with the modern treatment of entanglement/brick-wall divergences and gauge edge modes (\cite{SusskindUglum1994,SolodukhinLRR,DonnellyWall2015,Fursaev1995}). The generalized second law can be checked directly in our quasi-stationary setting (Sec.\,\ref{subsec:GSL-consistency}) and follows in full generality from the horizon-algebra framework (\eqref{eq:GSL-Wall}; \cite{Wall2012GSL,EngelhardtWall2015QES,BoussoEngelhardt2016GSL}).\\

Our leading-order statements assume (a) a static, spherically symmetric exterior; (b) a neutral scalar field; (c) constant Robin data; and (d) the ultrarelativistic regime underlying Secs.~\ref{sec:canonical} and \ref{sec:massless}. Low-$T_A$ features (e.g.\ canonical plateaus set by the lowest eigenvalues) depend on the intrinsic 3D geometry and on the Robin parameter $\tilde h$; the optimization framework in Sec.~\ref{sec:canonical} and App.~\ref{app:elliptic} makes this dependence precise. Beyond these caveats, the Weyl-controlled leading laws are insensitive to foliation and to the choice between Dirichlet/Neumann/constant Robin boundary data.

\paragraph{Outlook.}
We conclude with several concrete directions that appear both tractable and consequential:
\begin{enumerate}
	\item \textbf{Rotating/charged black holes.} Extending the micro--macro identity \eqref{eq:recall-AE} to Kerr and Reissner--Nordstr\"om backgrounds should incorporate the work terms from the first law \eqref{eq:threeone-firstlaw}. The expectation is that the leading Weyl-controlled exponents remain unchanged, while $\Omega_H$- and $\Phi_H$-dependent prefactors correct the analogues of \eqref{eq:state-eq-ultra} and \eqref{eq:sigma}.
	\item \textbf{Running horizon impedance.} Allowing the Robin parameter to depend on scale/state would probe how $a_1$-level data (App.~\ref{app:weyl-higher}) feed into observable offsets (cf.\ \eqref{eq:deltaA-explicit}) without altering the universal slopes fixed by $C_0$. This may connect to membrane-paradigm transport near the horizon (\cite{Solodukhin2001Bound,Park2023}).
	\item \textbf{Field content and statistics.} Fermions and gauge fields can be handled within the same spectral machinery; at low $T_A$ the canonical behavior differs (no macroscopic occupation), while at high $T_A$ the leading exponents again follow from $C_0$. Edge modes adjust the renormalization of $1/G$ but do not produce parametrically large finite terms (\cite{DonnellyWall2015,SolodukhinLRR}).
	\item \textbf{Dynamics beyond strict stationarity.} Adiabatic evaporation or accretion can be treated by promoting $E\!\to\!E(t)$ in \eqref{eq:recall-AE} and using instantaneous spectra; a fully dynamical formulation would profit from the ``entropy-maximization'' perspective on black-hole interior geometry \cite{Yokokura2025} together with the covariant GSL framework \cite{Wall2012GSL,EngelhardtWall2015QES}.
	\item \textbf{Numerics and spectral theory.} High-precision computation of the Heun roots entering \eqref{eq:heun-quantization} for general $\ell$ and $\tilde h$ would calibrate the low-$T_A$ regime and benchmark the asymptotics. Cross-checks across different horizon-regular slicings (App.~\ref{app:pg-slicing}) would quantify how prefactors vary while the exponents remain universal.
	\item \textbf{Information-theoretic synergies.} Because the framework elevates $A$ itself to the thermodynamic constraint, it meshes naturally with covariant entropy bounds \cite{Bousso1999} and near-horizon conformal enhancements \cite{CamblongOrdonez2007}. Exploring these links may help isolate which interior degrees of freedom are most relevant for the macroscopic area relations.
\end{enumerate}

In sum, the present scheme provides a concise, controllable bridge between interior spectral data and macroscopic area relations. Its predictive power comes from the spatial Weyl coefficient $C_0$, which fixes the exponents in the canonical and radiation sectors (Secs.~\ref{sec:canonical}--\ref{sec:massless}) and remains intact in four dimensions (Sec.~\ref{sec:4D-upgrade}). While quantitative refinements remain to be worked out, the core message is robust: once the area is taken as the thermodynamic constraint, the leading macroscopic laws follow from universal geometric spectral data.

\newpage
\appendix
\section{Appendix}

\subsection{Weyl coefficients beyond the $\lambda$--term (Robin)}
\label{app:weyl-higher}
For completeness we note the next terms in the Weyl/heat--kernel expansion for $-\Delta$ on a smooth three--manifold with boundary and Robin condition $\partial_{n}\psi + h\,\psi=0$ with constant $h\ge0$.
In terms of the heat kernel coefficients
\begin{equation}
	\Tr e^{-t\Delta} \sim (4\pi t)^{-3/2}\Bigl[a_{0} + a_{1/2}\, t^{1/2} + a_{1}\, t + O(t^{3/2})\Bigr],
\end{equation}
one has $a_{0}=\Vol_{h}(B_{\rs})$ and $a_{1/2}=\pm\sqrt{\pi}\,\Area_{h}(\partial B_{\rs})$ with the upper (lower) sign for Neumann/Robin (Dirichlet).
The first boundary curvature/impedance contributions enter via $a_{1}$ as
\begin{equation}
	a_{1}=\int_{\partial B_{\rs}}\!\Bigl(\alpha_{1} H + \alpha_{2} h + \alpha_{3} h^{2}\Bigr)\,dA \;+\; \int_{B_{\rs}}\!\alpha_{4}\,R\,dV,
\end{equation}
where $H$ is the mean curvature of $\partial B_{\rs}$, $R$ the scalar curvature of $h$, and $\alpha_{i}$ are dimensionless constants fixed by convention (see \cite{Vassilevich2003,Gilkey1995}).
Laplace inversion yields the counting function in the form
\begin{eqnarray}
	N(\lambda)&=&\frac{\Vol_{h}(B_{\rs})}{6\pi^{2}}\lambda^{3/2} + C_{1}(\mathrm{b.c.})\,\Area_{h}(\partial B_{\rs})\,\lambda \nonumber\\
	&+& c_{1/2}\,\biggl[\int_{\partial B_{\rs}}\!(H + \beta_{1} h + \beta_{2} h^{2})\,dA + \beta_{3} \int_{B_{\rs}}\! R\,dV\biggr] \lambda^{1/2} + O(\lambda^{0}),
\end{eqnarray}
with $C_{1}(\mathrm{Robin})=C_{1}(\mathrm{Neumann})=+1/(16\pi)$ and $C_{1}(\mathrm{Dirichlet})=-1/(16\pi)$, cf.~\eqref{eq:weyl-lambda}.
Thus a \emph{constant} Robin parameter $h$ does not affect the $\lambda$ term; its first effect appears at order $\lambda^{1/2}$ through the indicated integrals, in line with the main text.

\subsection{Alternative slicing: Painlev\'e--Gullstrand (PG)}
\label{app:pg-slicing}
As a second explicit check, consider a horizon-regular Painlev\'e--Gullstrand time function and take a constant-PG-time slice inside the horizon.
The induced spatial metric on such a slice is (globally) flat, so $\sqrt{\det h_{\mathrm{PG}}} = r^{2}\sin\vartheta$ and hence
\begin{equation}
	\Vol_{h}^{\mathrm{PG}}(B_{\rs})=4\pi\int_{0}^{\rs} r^{2}\,dr=\frac{4\pi}{3}\,\rs^{3},\qquad
	\Area_{h}(\partial B_{\rs})=4\pi \rs^{2}.
\end{equation}
Inserting this into~\eqref{eq:weyl-x} gives
\begin{equation}
	C_{0}^{\mathrm{PG}}=\frac{\Vol_{h}^{\mathrm{PG}}(B_{\rs})}{6\pi^{2}\rs^{3}}=\frac{2}{9\pi},\qquad
	C_{1}^{\mathrm{PG}}=\pm\frac{1}{4}\ \ \text{(Dirichlet / Neumann or Robin)}.
\end{equation}
Consequently, all leading \emph{exponents} in the areal thermodynamics (canonical $A\propto T_A$; radiation $A\propto T_A^{4}$) remain unchanged,
while \emph{prefactors} (e.g.\ the Stefan--Boltzmann--like constant $\sigma$ in~\eqref{eq:sigma}) rescale by $C_{0}^{\mathrm{PG}}/C_{0}=\tfrac{2}{9\pi}\big/\tfrac{5}{24}=\tfrac{16}{15\pi}$.

\subsection{Ellipticity, self-adjointness and boundary conditions}
\label{app:elliptic}
On the Riemannian manifold $(B_{\rs},h)$, the Laplace--Beltrami operator
\(
\Delta = h^{-1/2}\,\partial_i\big(h^{1/2} h^{ij}\partial_j\big)
\)
is elliptic with smooth coefficients. With the Robin boundary condition~\eqref{eq:robin-bc} the associated quadratic form is coercive and defines a unique self-adjoint realization on $L^2(B_{\rs},\sqrt{h}\,d^3x)$ with compact resolvent, hence purely discrete spectrum. Dirichlet and Neumann are recovered for $h\to\infty$ and $h=0$.

\subsection{Fluctuation--dissipation relation for the area operator}
\label{app:fluct}
From $\log Z_N=N\log z(\mu)-\log N!$ and $\mu=\gamma a$,
\begin{equation}
A=-\pdv{\log Z_N}{\gamma}= -N a\,\pdv{\log z}{\mu},\qquad
\pdv{A}{\gamma}=-\pdv[2]{\log Z_N}{\gamma}= -N a^2\,\pdv[2]{\log z}{\mu}.
\end{equation}
Since the variance is $\sigma_A^2=\pdv[2]{\log Z_N}{\gamma}$, we have $\sigma_A^2=-\pdv{A}{\gamma}$. Using $\gamma=1/(\kB T_A)$ then gives the standard identity
\begin{equation}
C\equiv \pdv{A}{T_A}=\pdv{A}{\gamma}\,\pdv{\gamma}{T_A}
=\frac{\sigma_A^2}{\kB T_A^2}\,,
\end{equation}
i.e. $\sigma_A^2=\kB T_A^2\,C$.

\end{document}